# Zinc adaptation and resistance to cadmium toxicity in mammalian cells. Molecular insight by proteomic analysis.

*Estelle Rousselet[1,2,3], Alain Martelli[1,2,3,4], Mireille Chevallet[3,5,6], Hélène Diemer[7], Alain Van Dorsselaer[7], Thierry Rabilloud[3,5,6] and Jean-Marc Moulis[1,2,3]*

[1] CEA, DSV, IRTSV, Laboratoire de Chimie et Biologie des Métaux, Grenoble, France.

[2] LCBM, CNRS, Grenoble, France.

[3] Université Joseph Fourier, Grenoble, France

[4] present address: IGBMC, 1 rue Laurent Fries, BP 10142, 67404 Illkirch, France

[5] CEA, DSV, IRTSV, Laboratoire de Biochimie et Biophysique des Systèmes Intégrés; Grenoble, France

[6] BBSI, CNRS, Grenoble, France

[7] Laboratoire de Spectrométrie de Masse, CNRS UMR7178, Strasbourg, France

Correspondence: Jean-Marc Moulis, CEA-Grenoble, IRSTV/LCBM, 17 rue des Martyrs, F-38054 Grenoble Cedex 9, France. Tel: (33) 4 38 78 56 23, e-mail: jean-marc.moulis@cea.fr

Abbreviations: HZR: High Zinc Resistant HeLa cells; SERCA: sarco-endoplasmic reticulum $Ca^{2+}$-ATPase; ICP-OES: Inductively coupled plasma optical emission spectroscopy; $LC_{50}$: lethal concentration for 50% of the cells; HSP: Heat Shock Proteins; UPR: Unfolded Protein Response; Hop: Hsp70/Hsp90 organizing protein; HPPD : 4-hydroxyphenylpyruvate dioxygenase; NTBC: 2-(2-Nitro-4-Trifluoromethyl-Benzoyl)-1,3-Cyclohexanedione; HGO : homogentisate 1,2-dioxygenase (EC E.C.1.13.11.5) ; ERAD: ER-associated protein degradation; MIF: macrophage migration inhibitory factor.

ABSTRACT

To identify proteins involved in cellular adaptive responses to zinc, a comparative proteome analysis between a previously developed high zinc- and cadmium- resistant human epithelial cell line (HZR) and the parental HeLa cells has been carried out. Differentially produced proteins included co-chaperones, proteins associated with oxido-reductase activities, and ubiquitin. Biochemical pathways to which these proteins belong were probed for their involvement in the resistance of both cell lines against cadmium toxicity. Among endoplasmic reticulum stressors, thapsigargin sensitized HZR cells, but not HeLa cells, to cadmium toxicity more acutely than tunicamycin, implying that these cells heavily relied on proper intracellular calcium distribution. The similar sensitivity of both HeLa and HZR cells to inhibitors of the proteasome, such as MG-132 or lactacystin, excluded improved proteasome activity as a mechanism associated with zinc adaptation of HZR cells. The enzyme 4-hydroxyphenylpyruvate dioxygenase was overproduced in HZR cells as compared to HeLa cells. It transforms 4-hydroxyphenylpyruvate to homogentisate in the second step of tyrosine catabolism. Inhibition of 4-hydroxyphenylpyruvate dioxygenase decreased the resistance of HZR cells against cadmium, but not that of HeLa cells, suggesting that adaptation to zinc overload and increased 4-hydroxyphenylpyruvate removal are linked in HZR cells.

**1 Introduction**

Zinc is an essential metal as a component of a wide variety of proteins, including transcription factors, metalloenzymes, and other proteins with various functions [1]. However, large or improperly handled zinc concentrations in cells are toxic [2]. Reciprocally, too little zinc induces apoptosis [3] since zinc is an inhibitor of caspase 3 [4]. Consequently zinc homeostasis is precisely maintained [5].

Zinc and cadmium share a range of physico-chemical properties, but it is not known whether their toxicity mechanisms are related or similar [6]. Cadmium is well recognized as an industrial pollutant and a food contaminant which accumulates with a half-life of more than 10 years in the body due to inefficient excretion mechanisms. Mercury is also a chemical sharing the same electronic configuration with filled *d* orbitals. However some properties of mercury, including very strong formation constants with various ligands, common two ligand-coordination, large volatility of the element, or ease of alkylation, trigger different toxicity and detoxification mechanisms in living cells and target different organs in mammals, such as the central nervous system, as compared to zinc and cadmium [7]. Cadmium is deleterious to HeLa cells, but adaptation to the toxic metal has been demonstrated [8]. A human epithelial cell line derived from HeLa cells by increased resistance to zinc, called HZR, has been previously developed [9] and it provides a convenient model to study mechanisms of resistance to heavy metals such as zinc and cadmium. This adapted cell line is maintained in 200 μM of zinc in the culture medium, a concentration that is more than 10 times the average zinc concentration (10-15 μM) found in human plasma [10].

Once inside cells, cadmium leads to a diverse and complex series of events that may culminate with cell death. To attenuate cadmium toxicity, cells may respond with various strategies. These include proteins and activities that: (i) sequester cadmium, such as metallothioneins (MT), to prevent its interaction with sensitive cellular targets; (ii) increase glutathione synthesis and detoxify reactive oxygen species that are generated by cadmium; (iii) repair damage to cellular components such as DNA; and (iv) help folding or degrade unfolded proteins [11]. In response to the endoplasmic reticulum (ER) stress that is induced by impairment of protein folding, cells activate a pathway known as the unfolded protein response (UPR) [12]. The UPR is a complex mechanism that includes increased synthesis of heat shock proteins (HSP) acting as mediators of protein folding to maintain or recover their activity. A mechanism that contributes to maintain cellular homeostasis is the elimination of unfolded proteins through the energy-dependent, ubiquitin-proteasome degradation pathway [13]. This pathway involves two steps. First, the target protein is conjugated with ubiquitin molecules at lysine residues, then the ubiquitin-tagged substrate is transferred to the 26S proteasome, a multisubunit complex consisting of a 20S barrel-shaped proteolytic core and a 19S cap-like regulatory complex.

A candid approach to gain insight into the differences between zinc-resistant HZR and reference HeLa cells has been implemented by obtaining high resolution 2-DE of total cellular extracts. Comparison of the two proteomes revealed a short list of proteins that differed between the two cell lines: they pointed to some of the above mentioned biological functions previously shown to be perturbed by cadmium. These functions have been probed by use of specific inhibitors and no unique activity can be singled out to explain the resistant phenotype of HZR cells. Rather, detailed analysis of the suggested pathways indicates a likely important role for signaling pathways, as shown by the sensitivity of resistant cells to perturbed calcium homeostasis. In addition, these studies also highlight the importance of 4-hydroxyphenylpyruvate dioxygenase, an enzyme of the tyrosine catabolic pathway, in the adaptation to zinc and the resistance against cadmium toxicity in HZR cells.

## 2 Materials and Methods

## 2.1 Cell Culture and Preparation of cell lysates

Epithelial human cervix carcinoma HeLa cells were grown in DMEM (Sigma-Aldrich, St Louis, MO, USA) with 2 mM L-glutamine and 5% heat-inactivated FBS at 37°C with a gas mixture of 5% $CO_2$ and 95% air. HZR cells [9] were routinely grown as HeLa cells in the same medium supplemented with 200 µM zinc sulfate. Exchange of zinc sulfate for zinc acetate did not change the behavior of HZR cells.

HeLa and HZR cells ($40 \times 10^6$) were harvested by centrifugation, rinsed three times in 1 mL phosphate-buffered saline and pellets were suspended in homogenization buffer (0.25 M sucrose, 10 mM Tris-HCl, pH 7.5, 1 mM EDTA). A buffer volume approximately equal to the packed cell volume was used. The suspension was transferred to a polyallomer ultracentrifuge tubes, and the cells were lysed by the addition of 4 volumes (respective to the suspension volume) of 8.75 M urea, 2.5 M thiourea, 25 mM spermine, and 50 mM dithiothreitol. After 1 hour at room temperature, the extracts were centrifuged (30 min at 200,000 x *g*). The supernatant was collected and the protein content was determined by the Bio-Rad protein assay using bovine serum albumin as a standard. A total of 500 µg of proteins were diluted in 1 mL rehydratation buffer (7 M urea, 2 M thiourea, 4% CHAPS, 0.4% ampholytes, 20 mM DTT). The protein extracts were stored at –20 °C.

## 2.2 2-DE

The data reported herein were obtained with samples prepared from at least three independent cultures of each cell line. For a given lysate, at least two gels were prepared, one to localize differentially produced spots between the two cell lines by silver staining and a second to confirm the first by visualizing with a mass spectrometry-compatible stain, and to extract proteins of interest.

The first dimension of electrophoresis was performed with immobilized pH gradients for isoelectric focusing. Non linear 4–8 and zoom 5.5-7.5 pH gradients were used. Homemade pH gradient plates were cast and cut into 4-mm-wide strips [14, 15]. The samples were applied onto the strips by in-gel rehydration overnight using a thiourea-urea mixture as denaturing agent [16]. IEF was carried out for 60,000 Vh at a maximum of 3000 V using the Multiphor II system (Amersham-Pharmacia, Sweden). Strips were then equilibrated for 20 min first in 0.15 M BisTris/0.1 M HCl, 6 M urea, 2.5% SDS, 30% glycerol, 0.5 M DTT and then in 0.15 M BisTris/0.1 M HCl, 6 M urea, 2.5% SDS, 30% glycerol, 0.3 M iodoacetamide. Strips were placed on top of a SDS-polyacrylamide gel. After migration, the gels were stained either with silver [17], or with colloidal Coomassie Blue when protein identification was sought [18]. Expression ratios were estimated by image analysis of pairs of gels with the Delta2D software (DECODON GmbH, Greifswald, Germany). At least three pairs (HeLa vs HZR cells) of gels run in parallel were analyzed and statistics reported in Table 1 are for between 3 and 6 measurements.

### 2.3 In Gel Digestion and MALDI-TOF-MS Analysis

Excising gel slices, rinsing, and reduction/alkylation steps were performed on a MassPREP station (Micromass, Manchester, UK) as described previously [19]. Gel pieces were completely vacuum-dried before digestion. Three volumes of freshly diluted 12.5 ng/µl trypsin (Promega, Madison, WI) in 25 mM $NH_4HCO_3$ were added to the volume of the dried gel. Digestion was performed at 35°C overnight. Then, the samples were again vacuum-dried for 5 min and 5 µl of 35% $H_2O$:60% acetonitrile:5% HCOOH were added to extract peptides. The mixture was sonicated for 5 min and centrifuged for 5 min. The supernatant was recovered and the procedure was repeated once with the pellet.

Mass measurements were carried out on an Ultraflex[TM] MALDI-TOF/TOF spectrometer (Bruker-Daltonik GmbH, Bremen, Germany). The instrument was used at a maximum accelerating potential of 20 kV and was operated in reflector-positive mode. Sample preparation was performed with the dried droplet method using a mixture of 0.5 µl of sample with 0.5 µl of matrix solution. The matrix solution was prepared from a saturated solution of α-cyano-4-hydroxycinnamic acid in $H_2O$:50% acetonitrile diluted 3 times in water. Internal calibration was performed with tryptic peptides resulting from autodigestion of trypsin (monoisotopic masses at $m/z = 842.5$; $m/z = 1045.6$; and $m/z = 2211.1$).

Monoisotopic peptide masses were assigned and used for databases searches using the search engine MASCOT (Matrix Science, London, UK) [20]. All human proteins present in Swiss-Prot were used without any p$I$ and $M_r$ restrictions. The peptide mass error was limited to 70 ppm, one possible missed cleavage was accepted.

**2.4 MTT cell viability and other assays**

The cytotoxicity of cadmium acetate, in addition to different drugs, on HeLa and HZR cells was determined after 24 hours of exposure by a modified (3-[4, 5-dimethylthiazol-2-yl]-2, 5-diphenyl tetrazolium) bromide (MTT) method [21]. Three thousand cells in 100 µl of growth medium were plated in 96-well plates and incubated for 24 hours before probing with a mix of cadmium acetate and different drugs. For HZR cells, 200 µM of zinc used to maintain these cells were replaced by cadmium. MTT, lactacystin, thapsigargin, and tunicamycin were obtained from Sigma Chemical Co (S[t] Louis, MO, USA). NTBC was a kind gift of Dr David King (Swedish Orphan International AB, Stockholm, Sweden): stock solutions were prepared in ethanol and diluted in the medium to a maximum of 0.1% ethanol. Viability was measured 24 hours later: ten µl of 5 mg/ml MTT were added in each well and the cells were incubated at 37°C for 3 hours. The medium was removed, and formazan was dissolved in 100 µl of DMSO. The optical density was measured at 560 nm with a microplate reader (Multiskan, Labsystem RC). The percentage of viable cells (%) was calculated as $[(A–B)/(C–B)] \times 100$ where $A = OD_{560}$ of the treated sample, $B = OD_{560}$ of the background absorbance, and $C = OD_{560}$ of the reference cells not exposed to the chemical compound being tested. Background absorbance was estimated by lysing a row of reference cells with 1% Triton X-100 before applying MTT. Error bars were calculated from 8 measurements in several separate experiments. The data were fitted to a sigmoid curve from which $LC_{50}$ values were derived as the 50% intersecting points. The statistical significance between data sets was assessed in Microsoft Excel by first comparing variances with a Fischer-Snedecor (F-) test for unpaired datasets at the 0.05 probability point. Student t-tests were then performed according to the outcome for F that generally indicated unequal variances.

For intracellular metal determinations, cells were harvested by trypsination and washed twice with PBS without calcium and magnesium and once with the same buffer containing 50 mM EDTA. Pellets were vacuum-dried and mineralized in 70% nitric acid before analysis with Inductively Coupled Plasma-Optical Emission Spectrometry (ICP-OES) with a Varian, Vista MPX instrument [22]. The metal content was referred to the amount of cells in the analyzed pellet.

Homogentisate dioxygenase was measured spectrophotometrically following formation of maleylacetoacetate at 330 nm, using an extinction coefficient of 13500 $M^{-1}$ $cm^{-1}$ [23]. First cell extracts were incubated anaerobically with 100 μM ferrous sulfate and 200 μM ascorbate for 10 min to reactivate any enzyme that may have been iron-depleted upon breaking the cells. The assay was carried out aerobically in 20 mM Mes, 80 mM NaCl at pH 6.2 with 2 mM homogentisate and it was started by adding the enzyme.

The intracellular melanin concentration was measured by a previously devised method [24] with ca. $5x10^{6}$ cell-samples and a reference plot obtained with synthetic melanin prepared by oxidation of tyrosine with hydrogen peroxide (Sigma, reference M8631) and dissolved in 0.85 N KOH. Briefly, cells were washed with 0.02% EDTA and lysed with 0.5 ml water followed by two cycles of freezing and thawing. After centrifugation, pellets were washed three times with 5% trichloroacetic acid, twice with a cold mixture of ethanol/ethyl ether (3/1) and once with cold ether. The air-dried pellets were dissolved in 0.3 ml of 0.85 N KOH by heating to 100°C for 10 min. The absorbance was measured at 400 nm.

## 2.5 Quantitative RT-PCR

RNA extractions were carried out with the total Quick RNA$^{TM}$ kit (Talent s.r.l., Trieste, Italy). Total RNA (1.4 µg) was reverse transcribed by the RevertAid$^{TM}$ H Minus M-MuLV Reverse Transcriptase (MBI Fermentas, Vilnius, Lithuania) at 42°C for one hour using a poly-T primer. Quantitative RT-PCR was performed with a Stratagene Mx3005P instrument as follows: the reaction mixture consisted of 12.5 µl of Full Velocity SYBR Green QPCR Master Mix (Stratagene, La Jolla, CA, USA), 140 nM of HPPD primers and 5 µl of 12.5 ng cDNA adjusted to 25 µl with water. Initial denaturation was at 95°C for 5 min, followed by amplification for 40 PCR-cycles at 95°C for 10 s and at 60°C for 30 s successively. PCR fluorescent signals were normalized to the fluorescent signal obtained from the housekeeping genes ribosomal phosphoprotein PO (RPLPo2), β-actin and glyceraldehyde 3-phosphate dehydrogenase (GAPDH) for each sample.

## 3. Results

### 3.1 Comparative Proteome analysis of HeLa and HZR cells.

To probe the mechanisms responsible for zinc resistance of HZR cells, 2-DE has been implemented to compare the HZR proteome with that of parental HeLa cells (Figure 1). Lysates of four different cultures of each cell line were analyzed. The general pattern displayed by all gels is quite similar, despite the addition of 200 μM zinc to the medium used to grow HZR cells (Figure 1). This indicates that such gels are highly reproducible and that only few proteins have changed intensities or positions between the two cell lines. Therefore these differences are probably not due to a general and massive effect of zinc overload in HZR cells, and they are likely to give insight into the mechanism responsible for the phenotype of HZR cells. More detailed data were obtained with p$I$ gradients between 5.5 and 7.5. Figure 2 shows silver-staining of representative gels obtained after separation of 500 μg protein samples from HeLa and HZR cells, respectively. Spots of different intensities or absent in one of the two gels were excised and subjected to trypsin digestion, MALDI-TOF mass spectra measurements, and database searching. Table 1 summarizes the proteins, the concentrations of which change between HeLa and HZR cells. This list includes a co-chaperone, stress induced phosphoprotein-1 or Hsp70/Hsp90 organizing protein (Hop), enzymes or subunits involved in oxidoreductase activities, thioredoxin reductase, flavoprotein subunit of succinate dehydrogenase, 4-hydroxyphenyl-pyruvate dioxygenase, and carbonic anhydrase II.

As Chimienti *et al*. [9] previously pointed out, metallothioneins largely contribute to zinc adaptation of HZR cells. In an attempt to visualize metallothionein up-regulation by 2-DE, higher concentration gels were applied in the second dimension. Two spots of apparent 6.5 kDa mass were detected. The protein labeled M1 is of higher intensity in gels of HZR extracts than in those of HeLa cells (Figure 3) but its p$I$ is lower than expected for metallothionein.

Indeed, identification with mass spectrometry revealed that M1 did not correspond to metallothionein but to ubiquitin. M2 is the ubiquitin-like molecule NEDD8, the function of which includes regulation of the cell cycle, but its intensity is weak and it does not vary between the two cell lines under study (Figure 3).

The role of the proteins listed in Table 1 in the cellular response to divalent metal stress has been probed in the following.

## 3.2 ER stress and sensitivity to cadmium toxicity

Stress-induced phosphoprotein 1 (Hop) is a 60-kDa co-chaperone that mediates the association of the heat shock proteins Hsp70 and Hsp90 [25]. The function of Hop has been best characterized in *in vitro* systems examining the assembly of the progesterone receptor [26] and glucocorticoid receptor [27] into hormone-binding competent hetero-complexes [28]. In previous investigations, cadmium has been reported to interfere with protein folding, leading to accumulation of misfolded proteins in ER [29], and HZR cells have been shown to sustain cadmium concentrations almost as high as zinc ones over 24 hours without increased lethality (Rousselet et al. in preparation). To estimate the endoplasmic reticulum stress that cells studied herein may experience under cadmium exposure, two different drugs were applied in addition to the toxic metal. The sesquiterpene lactone, thapsigargin, inhibits sarco-endoplasmic reticulum $Ca^{2+}$-ATPase (SERCA) and consequently depletes the intracellular calcium stores [30] and tunicamycin inhibits N-acetylglucosamine transferases, so preventing glycosylation of newly synthesized glycoproteins.

Viability tests were performed with the MTT assay in the presence of cadmium acetate for HeLa and HZR cells with 50 nM of thapsigargin or tunicamycin. These concentrations of inhibitors were chosen because they minimally decrease viability to the same extent (less than ca. 20%) for both cell lines. After 24 hours of exposure, the half-maximal lethal concentrations were about 7 µM of cadmium for HeLa cells exposed to 50 nM thapsigargin or 50 nM tunicamycin (Figure 4A). These values are not significantly different than those measured in the absence of thapsigargin or tunicamycin (Rousselet et al. in preparation), indicating that these concentrations of ER stressors do not sensitize HeLa cells to the cadmium insult.

In the case of HZR cells, replacement of the 200 µM zinc sulfate by varying cadmium concentrations for 24 hours and using the same concentration of thapsigargin as for HeLa cells indicated that these cells were far more sensitive to cadmium than without thapsigargin (Figure 4B). Cadmium half-maximal lethal dose measured after 24 hours of exposure was 25 µM with 50 nM thapsigargin. This value is ten times smaller than in the absence of thapsigargin. Exposure of HZR cells to 50 nM tunicamycin (Figure 4B) led to a cadmium half-maximal lethal dose of 165 µM, but this increased sensitivity to cadmium with tunicamycin is not as large as with thapsigargin. Therefore, inhibition of SERCA by thapsigargin and subsequent disruption of calcium homeostasis appears as a more potent stress that enhances cadmium toxicity than inhibition of glycosylation. These experiments indicate that increasing ER stress decreases cadmium resistance toward cadmium, in a very sensitive way for HZR cells treated with thapsigargin.

Since HZR cells showed a strong sensitivity to the presence of 50 nM thapsigargin, the metal content of cells under the conditions implemented to assess viability in the presence of cadmium was measured by ICP-OES. Thapsigargin lowered cadmium accumulation in HeLa cells, but not in HZR cells, at least in the cadmium concentration range that remained accessible with 50 nM of the drug. It follows that the loss of viability of HZR cells is not directly linked to intra-cellular cadmium in the presence of thapsigargin.

### 3.3 Proteasomal degradation and sensitivity to cadmium toxicity

As shown in Figure 3, ubiquitin was found to be up-regulated in HZR cells maintained in 200 µM zinc. Although ubiquitin fulfills widespread cellular functions [31], labeling of misfolded proteins for removal by the proteasome is one of the major role of this small protein [13]. Up-regulation of ubiquitin in HZR cells under standard conditions, i.e. with 200 µM zinc, may reveal enhanced protein turnover through the proteasome pathway.

This pathway was probed by viability tests implementing the MTT assay on HeLa and HZR cells with increasing concentrations of lactacystin [32] for 24 hours (Figure 5A). A proteasomal inhibitor concentration (2.5 µM) leaving about 80% of the cells under study viable after 24 hours of exposure was chosen in the following experiments. Half-maximal lethal concentration after 24 hour-exposure to cadmium and 2.5 µM lactacystin was 3 µM for HeLa cells (Figure 5B), as compared to 7 µM for cells exposed to cadmium alone ($p$ value <0.001). In the case of HZR cells, the cadmium half-maximal lethal dose was 75 µM after 24 hours of exposure to 2.5 µM lactacystine (Figure 5C). As in the case of HeLa cells, this value is significantly smaller ($p$ value <0.001) than that measured in the absence of lactacystin. Qualitatively similar results were obtained by replacing the proteasomal irreversible inhibitor lactacystin by the reversible inhibitor MG132. These data highlight the involvement of the ubiquitin-proteasome pathway in the cellular response against cadmium toxicity in both HeLa and HZR cells.

**3.4 Involvement of 4-hydroxyphenylpyruvate dioxygenase activity in zinc adaptation and cadmium toxicity for HZR cells**

The 2-DE data revealed up-regulation of 4-hydroxyphenylpyruvate dioxygenase (HPPD) in HZR cells as compared to HeLa cells (Figures 1 and 2). We have thus evaluated expression of the HPPD gene in both cell lines: relative levels of HPPD transcripts were measured by RT-PCR under standard growth conditions and compared to three different housekeeping genes. The expression of the HPPD gene was found to be about five times higher in HZR than in HeLa cells (Figure 6). This value agrees with the concentration ratio estimated for the corresponding proteins (Table 1). Therefore HPPD overproduction in HZR cells occurs mainly by increased transcription or mRNA stabilization as compared to HeLa cells. The increased concentration of HPPD in HZR cells does not seem to be triggered by the high zinc concentration in which these cells are maintained. Indeed, HeLa cells exposed to zinc before analysis by 2-DE did not display higher HPPD concentrations than HeLa cells kept in a medium with a normal zinc concentration. However, the reversibility of the HZR phenotype [4] was mirrored by the decrease of HPPD in HZR cells kept without excess of zinc for 3 days. Therefore, increased HPPD is associated with the HZR phenotype characterized by resistance to both high zinc and cadmium concentrations, rather than with up-regulation induced by short-term exposure to zinc.

In order to know if the HPPD activity contributes to cadmium management in the presently studied cells, a specific inhibitor of the enzyme, 2-nitro-4-(trifluoromethyl) benzoyl-1,3-cyclohexanedione (NTBC), was used [33, 34]. HeLa and HZR cells were treated with different concentrations of NTBC for 24 hours. Up to 300 µM of NTBC, viability slowly decreased in a similar dose-dependent way for both cell lines (Figure 7A). The NTBC concentration of 5 µM minimally affected viability. It was added with cadmium acetate for 24 hours to measure the effect of the inhibitor on the cellular sensitivity to cadmium. For HeLa cells (Figure 7B), the half-maximal lethal cadmium concentration was unchanged in the presence of NTBC as compared to HeLa cells without NTBC ($p$ value between data sets with and without the inhibitor= 0.63), whereas $LC_{50}$ was 90 µM of cadmium in the presence of 5 µM NTBC for HZR cells (Figure 7C). No significant changes of the intra-cellular concentrations of cadmium were measured for cells exposed to the same extra-cellular cadmium concentration, whether 5 µM of NTBC were added or not. Therefore HeLa cells did not seem to be more sensitive to cadmium in the presence of NTBC, but viability of HZR cells decreased by about three fold in these conditions (Figure 7C). The $LC_{50}$ shifts for HZR cells in the presence of NTBC varied with the concentration of the inhibitor, thus showing that NTBC sensitizes HZR cells for cadmium insult in a dose-dependent way.

### 3.5 Sensitivity of HZR cells to combined HPPD inhibition and cadmium treatments after long-term zinc removal

The above experiments were carried out by cadmium substitution of zinc in the HZR growth medium for 24 hours. However the zinc and cadmium resistance phenotype of HZR cells is reversible upon zinc withdrawal [9]. When zinc is removed from the growth medium for one week, not only the resistance against zinc decreases [9], but also the resistance against cadmium collapses to reach a $LC_{50}$ of 40 µM for 24 hours of exposure to cadmium (Rousselet et al. in preparation). These long-term zinc-depleted HZR cells were also treated with 5 µM or 10 µM NTBC in addition to cadmium for 24 hours. Viabilities obtained were similar for both NTBC concentrations giving $LC_{50}$ values of 40 µM on average. Therefore resistance against cadmium toxicity of long-term zinc-depleted HZR cells was greatly decreased as compared to zinc-maintained HZR cells, and NTBC no longer had any effect on the half-maximal lethal cadmium concentration for these cells, as was observed for HeLa cells (Figure 7B). These experiments demonstrate that cadmium handling by HZR cells depends on HPPD activity. Cadmium sensitivity increases when this activity is inhibited (Figure 7) and phenotypic reversal by long-term zinc removal makes these cells insensitive to HPPD inhibition.

### 3.6 Combined effects of added tyrosine and HPPD inhibition on cadmium toxicity

The tyrosine degradation product, 4-hydroxyphenylpyruvate (HPP), is converted by HPPD to homogentisate [35]. Since HZR cells over-produce HPPD, tyrosine catabolism may be more active in these cells than in HeLa cells. Addition of tyrosine for 24 hours is not toxic for both cell lines up to 6 mM (not shown). The growth medium of HeLa and HZR cells, which already contains about 230 µM of this amino acid, was supplemented with sub-lethal 3 mM of L-tyrosine. Viability curves were recorded for HeLa and HZR cells under these conditions with increasing cadmium concentrations. They show a decrease of $LC_{50}$ with 3 mM L-tyrosine, which goes down to 5 µM for HeLa cells (Figure 8A) and 145 µM for HZR cells (Figure 8B), instead of 7 µM and 250 µM, respectively, for cells exposed to cadmium alone (Figure 8). Therefore, the burden of cadmium challenge is increased by the presence of sub-lethal tyrosine concentrations for both cell lines. The combined effects of NTBC and tyrosine on the sensitivity toward increasing cadmium concentration were also determined. The cadmium $LC_{50}$ obtained for HeLa cells in the presence of 3 mM tyrosine and 5 or 10 µM NTBC did not change as compared to tyrosine alone. In the case of HZR cells, the same experiments with both NTBC and tyrosine gave the same outcome, namely no significant changes of $LC_{50}$ for cadmium as compared to cells exposed to tyrosine alone.

The enzyme following HPPD in the tyrosine degradation pathway is homogentisate dioxygenase (HGO), that is mainly responsible for homogentisate withdrawal in mammalian cells. The activity of this enzyme in extracts of both HeLa and HZR cells has been measured. As found for HPPD, HGO activity is ca. 5-fold higher in HZR cells (27 (SD 6) $µM.min^{-1}.(mg$ of proteins$)^{-1}$), as compared to HeLa cells (5.5 (SD 3.6) $µM.min^{-1}.(mg$ of proteins$)^{-1}$). Therefore, 4-hydroxyphenylpyruvate conversion and ring opening are both enhanced in HZR cells as a result of adaptation to zinc.

Tyrosine has different biological roles in mammalian cells, as one of the component of proteins, but also as the precursor of catecholamines, such as dopamine, thyroid hormones, and melanin. The latter pigment has been shown to bind metals [36, 37], including zinc and cadmium, and it has been repeatedly proposed to participate in heavy metal detoxification [38, 39]. Homogentisate is a precursor of secreted melanin pigments (pyomelanin) in some microorganisms [40] and of plasma melanins in humans in cases of alkaptonuria with inflammation and darkening of connective tissues in particular [41]. It is unknown at present if homogentisate, or a derivative, has the ability to produce heavy metal-protecting scavengers in mammalian cells. However, it is unlikely that these scavengers are in the form of intracellular melanin, since the measured amount in HZR cells (range 21-110 ng/µg of total proteins, 4 measurements) is not larger than in HeLa cells (range 45-180 ng/µg of total proteins, 3 measurements). Therefore the increased activities of HPPD and HGO in HZR cells do not favor melanin biosynthesis.

## 4. Discussion

In the present study adaptation to zinc overload developed by the HZR clone [9] was probed at the proteomic level. A list of proteins was found to be up-regulated in these cells, with the exception of carbonic anhydrase II that was decreased by a factor of 3 (Table 1). The corresponding gene was shown to be up-regulated by zinc deficiency in rat esophageal epithelia [42], in line with down-regulation upon zinc overload as noticed here. The biological pathways and functions suggested by comparative studies between the HZR clone and parental HeLa cells have been interrogated by use of specific inhibitors. Consequences on the sensitivity to cadmium have been measured to gauge the involvement of these pathways in the toxicity mechanism(s).

ER as a target of cadmium action in mammalian cells has been largely documented [43, 44]. The viability of HZR cells was far more sensitive than that of HeLa cells to the simultaneous addition of the ER stressors tunicamycin or thapsigargin with cadmium. This strongly suggests that the HZR cells, able to accommodate high intracellular zinc concentrations [9], more heavily rely on ER functions than HeLa cells. However, ER-associated chaperones, such as Grp94, PDI, and Grp78, or other components of the quality control system of protein folding, such as Hsp90, do not show significant changes between the two cell lines (Figure 1). It means that HZR cells do not display signs of generalized misfolding of proteins as compared to HeLa cells before application of cadmium.

Once cadmium is added to the cultures, perturbed protein biosynthesis does occur as witnessed by the impairment of resistance against the toxic metal by small amounts of proteasomal inhibitors. But, under these conditions, HeLa cells appear as sensitive as HZR cells, since the decrease of viability is similar for the same inhibitor concentration. The applied cadmium concentrations are quite different between the two cell lines, but it has been found that viability and the intracellular concentrations of the metal are similarly related for both HeLa and HZR cells (Rousselet et al. in preparation). Therefore, the increased amount of ubiquitin in HZR cells does not seem to be needed for a more efficient degradation of misfolded proteins via the proteasome pathway. It may rather be due to some other role of ubiquitin, such as protein intracellular trafficking and other important biological functions [45].

The only detected exception to the similar concentration of chaperones between the two cell lines (Table 1) is Hop that is more abundant in HZR cells than in HeLa cells. Hop belongs to the large group of co-chaperones which regulates the activity of heat shock proteins through its tetratricopeptide repeat TPR1 and TPR2A domains [46]. It participates in the maturation of the glucocorticoid receptor [47] and up-regulation of Hop has been demonstrated in activated macrophages [48].

This single example among the several chaperones identified in 2-DE strongly suggests that adaptation of HZR cells to zinc is a well-targeted phenomenon involving only a small number of cellular components. Further experimental support for this statement is provided by the more acute sensitivity of these cells to the combined effects of cadmium and thapsigargin as compared to cadmium and tunicamycin. These two drugs target distinct activities, but they eventually trigger the same cellular Unfolded Protein Response. The larger effect of thapsigargin than that of tunicamycin in the presence of cadmium indicates that toxicity of the metal in HZR cells is more sensitive to adequate calcium intracellular distribution than to proper processing of membrane and secreted proteins.

Besides Hop, another protein being unambiguously present at higher concentration in HZR cells than in HeLa cells is the enzyme HPPD. It is an iron- and 2-oxo-glutarate-dependent enzyme that catalyzes the complex conversion, including hydroxylation, decarboxylation and rearrangement steps, of 4-hydroxyphenylpyruvate, the product of tyrosine aminotransferase, to homogentisate [49, 50].

Missense mutations in the HPPD locus can cause two distinct genetic diseases, hereditary type III Tyrosinemia and Hawkinsinuria [51]. Type I Tyrosinemia is due to fumarylacetoacetate hydrolase deficiency [52], and, because symptoms of type III Tyrosinemia and Hawkinsinuria [53] are less severe than those of type I Tyrosinemia, the latter is treated by inhibition of HPPD with the 1,3 diketone NTBC [34], hence avoiding the buildup of the deleterious metabolites fumarylacetoacetate and succinylacetone. These metabolic and clinical data clearly indicate that defaults of tyrosine catabolism strongly impact cellular fate.

HPPD is subjected to several post-translational processing events [35, 54], which may explain the presence of at least two spots on 2D gels assigned to HPPD (Figure 2, Table 1), but no variations between the relative intensities of these spots have been observed in our experiments. The HPPD transcript is more abundant in the HZR than in the parental HeLa cell line (Figure 6). Promoter analysis evidenced binding sites for several transcription factors, including Sp1, CREB, and C/EBP [54]. The HPPD activity appeared more important in HZR cells than in HeLa cells, since its inhibition selectively sensitized the former cells to cadmium. However there is no association between HPPD activity and cadmium traffic since NTBC did not change the amount of metal accumulated by cells. A modest impact of increased tyrosine on cadmium toxicity was measured. If tyrosine removal has a role in cadmium detoxification, inactivation of HPPD should enhance cadmium toxicity in the presence of large amounts of tyrosine. Instead, no additive effects of tyrosine and the HPPD inhibitor NTBC on the viability curves with cadmium were detected. The NTBC dose-dependent decrease of HZR cells viability in the presence of cadmium is thus not due to decreased tyrosine disposal.

The HPPD substrate, HPP, is the first compound produced in the tyrosine degradation pathway. It has also been shown to be a substrate of phenylpyruvate tautomerase, which produces the enol-form of HPP and is the same protein as macrophage migration inhibitory factor (MIF) [55]. Whether increased depletion of HPP, as expected in HZR cells, influences the cytokine, including growth-promoting [56], function of MIF is unknown. Indeed, regulation of the balance between the two activities of this moonlighting protein is not elucidated, and the physiological meaning of the enzymatic reaction has been questioned [57]. Yet, from what is known about regulation of enzymes involved in tyrosine turnover, including HPPD [54], it is beyond doubt that signaling pathways regulate flow between the different metabolic routes available for tyrosine. In this respect, disruption of calcium homeostasis by thapsigargin in HZR cells, and its consequences on signaling pathways, strongly impacts the resistance of these cells against cadmium (Figure 4B). MIF may also be involved in regulating signaling pathways [58].

The proteomic approach described herein and complementary experiments have revealed an unsuspected association between tyrosine catabolism and adaptation to zinc overload that also contributes to the toxicity of cadmium in mammalian cells. Very recently [59], up-regulation of HPPD has been observed in *Arabidopsis thaliana* upon exposure to cadmium, as a means to increase vitamin E synthesis for which homogentisate is a precursor in plants. Since this pathway is not present in animals, the association between HPPD activity and cadmium toxicity in different organisms will be worth investigating further.

**Acknowledgements**

This work was supported by a grant from the "*Toxicologie Nucléaire et Environnementale*" program. Dr David King (Swedish Orphan International AB, Stockholm, Sweden) is gratefully thanked for providing us with NTBC, as is Dr Pierre Richaud (CEA, IBEB, LB3M, Cadarache, France) for ICP-OES measurements.

**Table 1. Identification of proteins differently present in HeLa and HZR cells[a].**

| Spot | Protein | Accession | MW / p$I$ | %R | E (SE) |
|------|---------|-----------|-----------|-----|--------|
| I | 4-Hydroxyphenylpyruvate dioxygenase | P32754 | 44775 / 6.50 | 44% | 4.7 (1.5) |
| II | Thioredoxin reductase GRIM-12 | Q9UES8 | 54580 / 6.36 | 21% | *b* |
| III | Stress-induced-Phosphoprotein 1 | P31948 | 62599 / 6.40 | 8% | 1.6 (0.3) |
| IV | Succinate dehydrogenase flavoprotein subunit, mitochondrial precursor | P31040 | 72645 / 7.06 | 29% | 2.8 (0.3) |
| V | Carbonic anhydrase II | P00918 | 29097 / 6.86 | 44% | 0.29 (0.09) |
| M1 | Ubiquitin | Q9VZL4 | 8560 / 6.56 | 85% | *c* |

[a] The labeled spots in Figure 2 are indicated with their identification using MALDI-TOF with sequence coverage (% R). Swiss-Prot accession numbers, monomer masses and isoelectro-focusing points obtained for each protein are also given. The average expression ratio between HZR and HeLa cells estimated with the Delta2D software are indicated under E with standard errors in brackets as described in Materials and Methods.

[b] the protein was not detected in HeLa cells

[c] the gels such as that in Figure 3 were not analyzed with the Delta2D software.

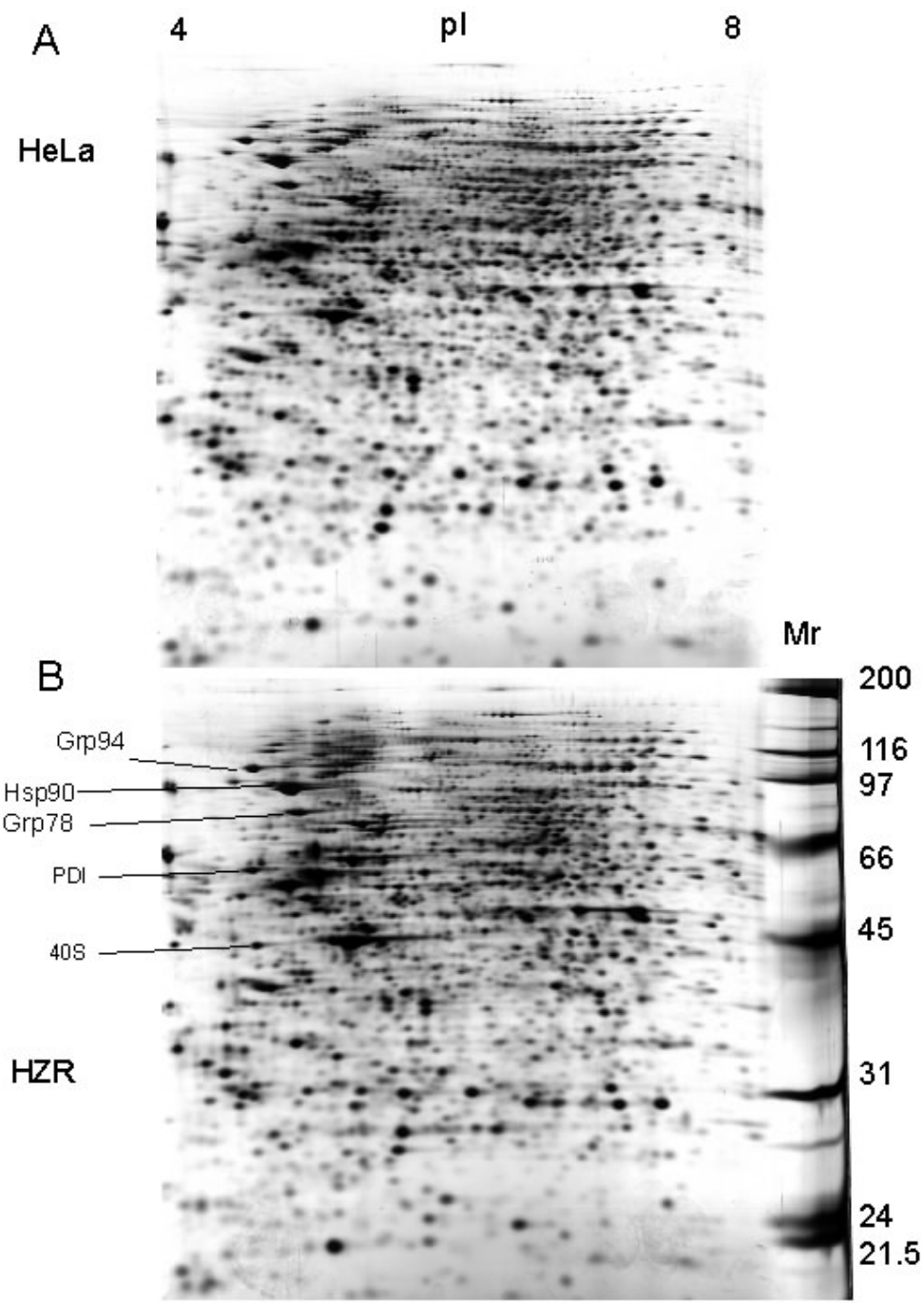

**Figure 1. 2-DE of total cell extracts from HeLa (A) and HZR (B) cells.** Two hundred micrograms of proteins were separated on pH 4-8 linear immobilized pH gradients (horizontal axis) and 10% continuous SDS gels (vertical axis). Molecular markers positions are indicated at the right of the gel in (B). Detection was by silver staining. Some easily identified proteins, mainly chaperones, are indicated in the bottom gel with PDI: protein disulfide isomerase and 40S: 40S ribosomal protein.

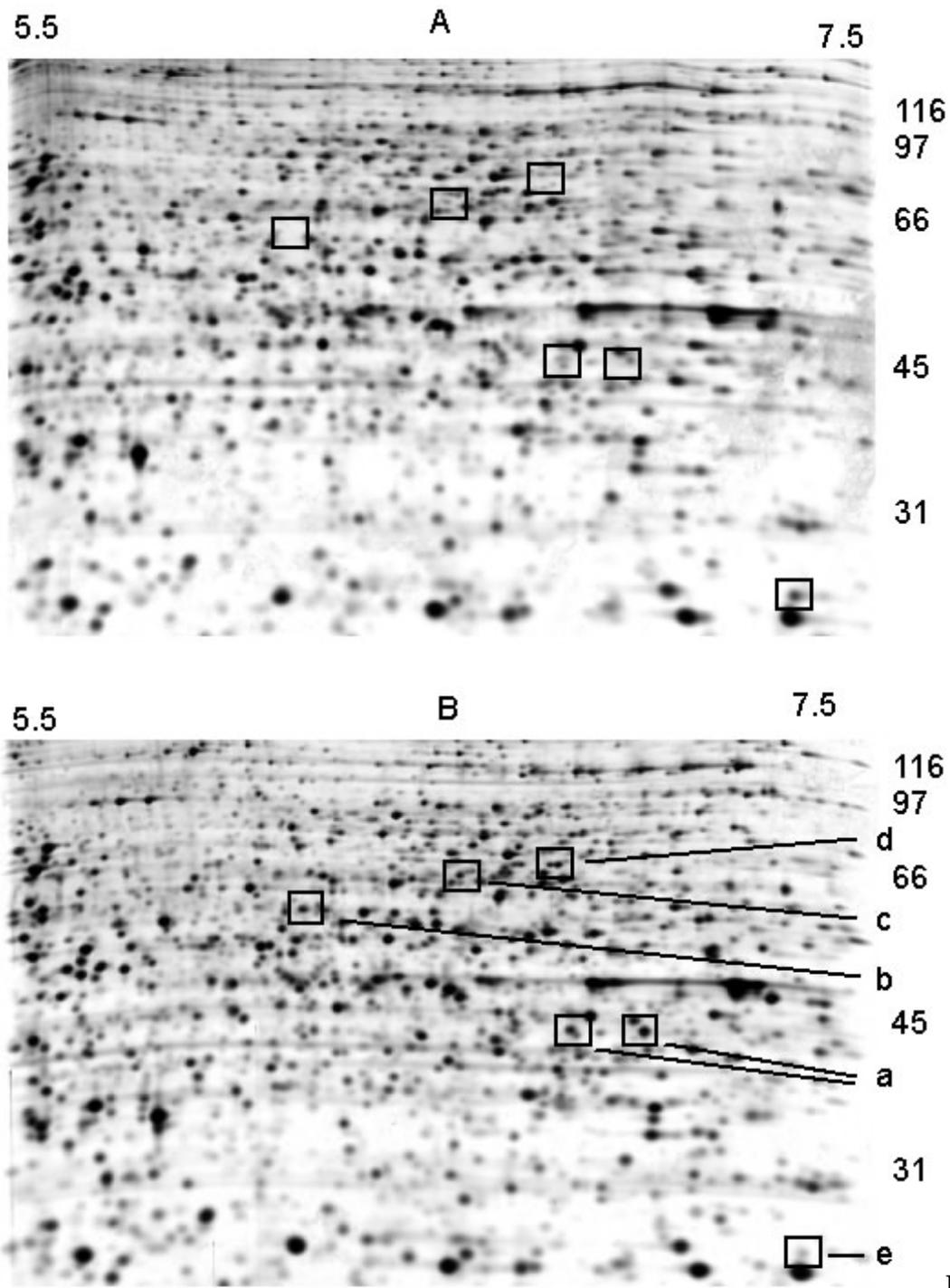

**Figure 2**.

**Highly resolved part of the 2-DE gels of total HeLa (A) and HZR (B) cell extracts.** Five

hundred micrograms of proteins were separated on pH 5.5-7.5 linear immobilized pH

gradients and 10% continuous SDS gels. Detection was by Coomassie Brillant Blue for micropreparative gels. Labeled spots were identified as reported in Table 1.

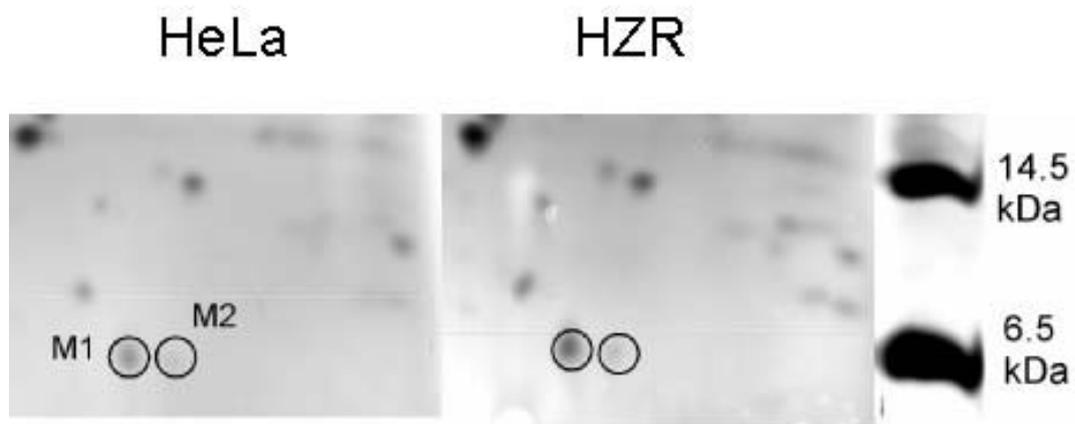

**Figure 3. Low molecular weight part of 2-D electrophoretic gels of total HeLa (left) and HZR (right) cell extracts**. Proteins were separated on pH 4-8 linear immobilized pH gradients and 12.5% SDS gels. The M1 and M2 spots are discussed in the text.

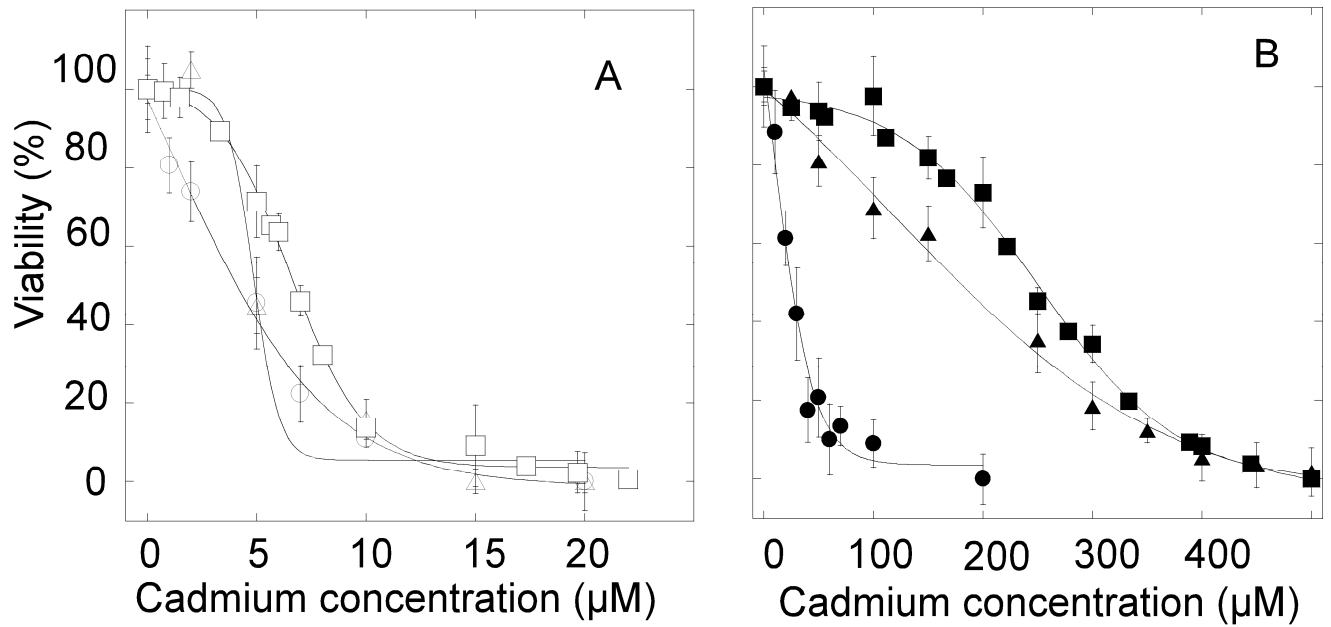

**Figure 4**. **Effects of ER stress on the viability of HeLa and HZR cells in the presence of cadmium.** HeLa cells were stressed by cadmium alone (**A** open square), in the presence of 50 nM thapsigargin (**A** open circle), or in the presence of 50 nM tunicamycin (**A** open triangle), 24 hours before measuring viability. The same experiments were done for HZR cells (**B**) and represented with the same filled symbols.

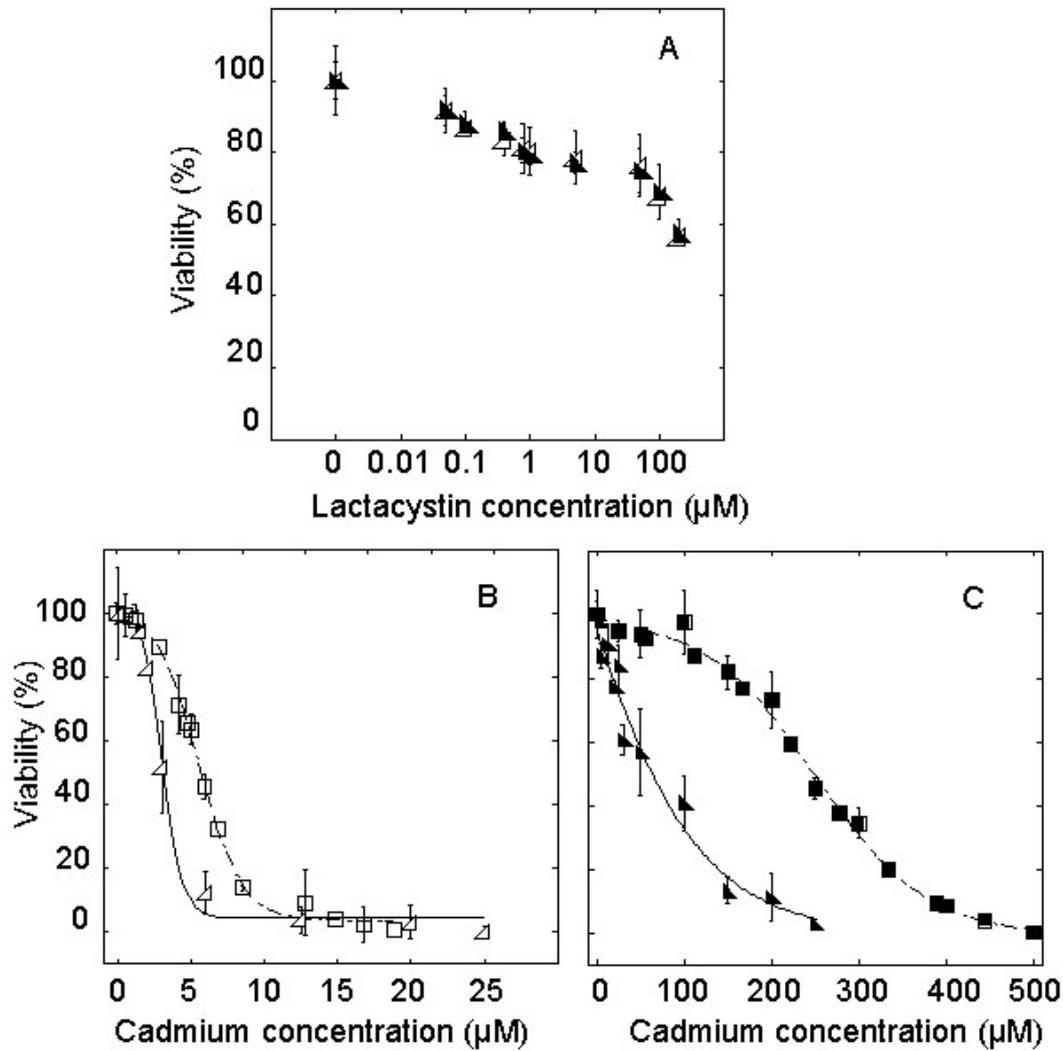

**Figure 5**. **Proteasome involvement in the sensitivity of HeLa and HZR cells to cadmium toxicity.** (**A**) Parental HeLa (open triangle) and HZR cells (filled triangle) were treated with increasing concentrations of lactacystin in the growth medium for 24 hours and the percentage of viable cells was measured and reported to that of non-exposed cells. (**B**) HeLa cells were stressed by cadmium alone (open squares) or in the presence of 2.5 µM lactacystin (open triangles) for 24 hours before measuring viability. (**C**) The same experiment as in (B) was carried out with HZR cells without (filled squares) or with 2.5 µM lactacystin (filled triangles).

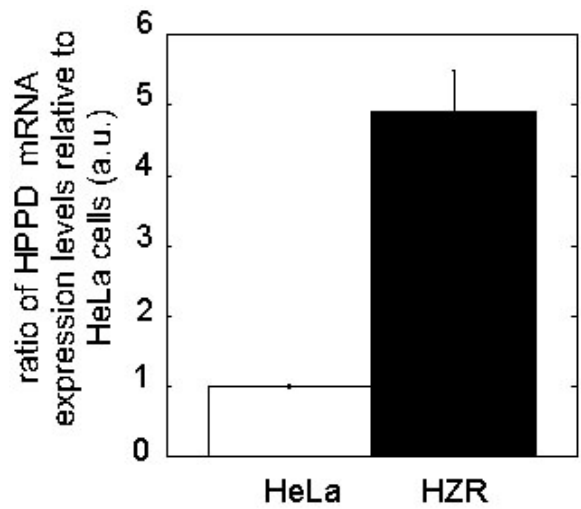

**Figure 6**. **Real-time PCR of HPPD.** Real-time quantitative, reverse transcriptase polymerase chain reaction experiments were carried out with total RNA purified from HeLa (empty bars) and HZR cells (filled bars). The results were calculated as a ratio to the set of reference genes (RPLPo2, GAPDH and β-actin) and the value for HeLa cells was arbitrarily set to 1. Data are the mean and SD of two separate experiments.

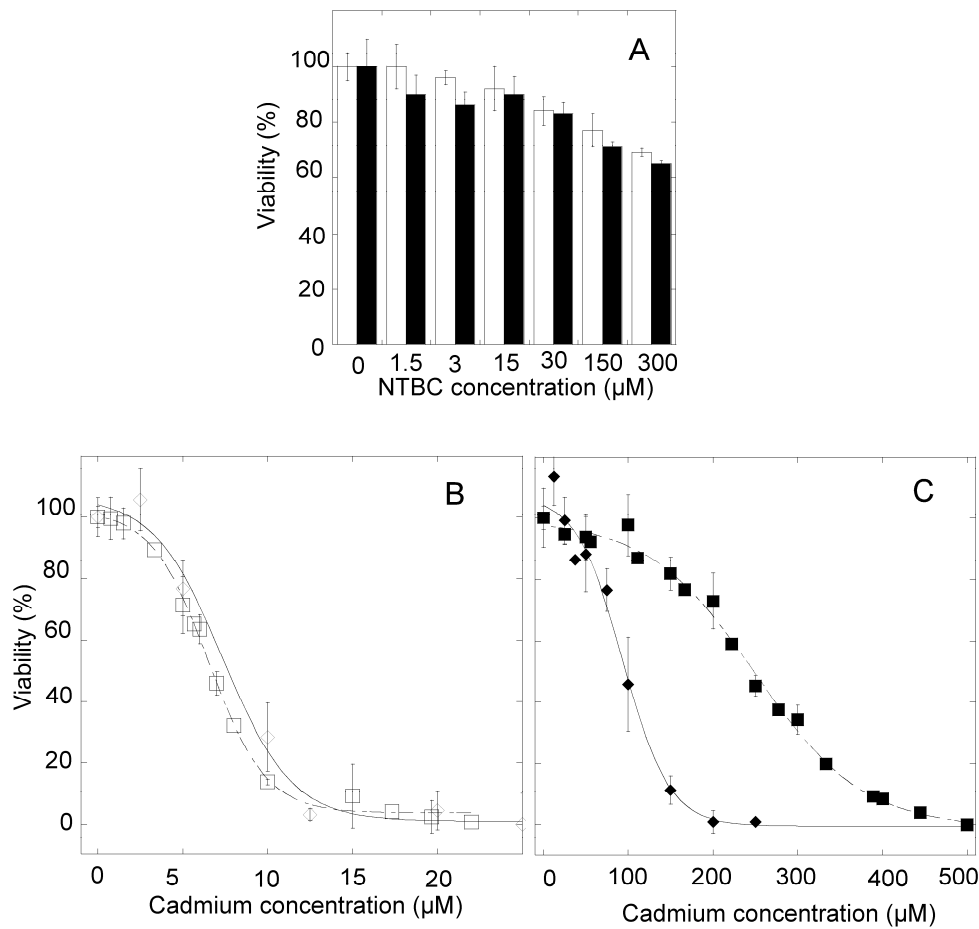

**Figure 7**. **HPPD inhibition and cadmium toxicity in HeLa and HZR cells. (A)** Parental

HeLa (open bars) and HZR cells (filled bars) were treated with increasing concentrations of

NTBC in the growth medium for 24 hours and the viability was estimated by the MTT assay.

in 8 separate measurements. **(B)** Parental HeLa cells were stressed by cadmium alone (open squares) or in the presence of 5 µM NTBC (open diamonds) 24 hours before measuring viability. **(C)** The same experiments were carried out with HZR cells and represented with the same filled symbols.

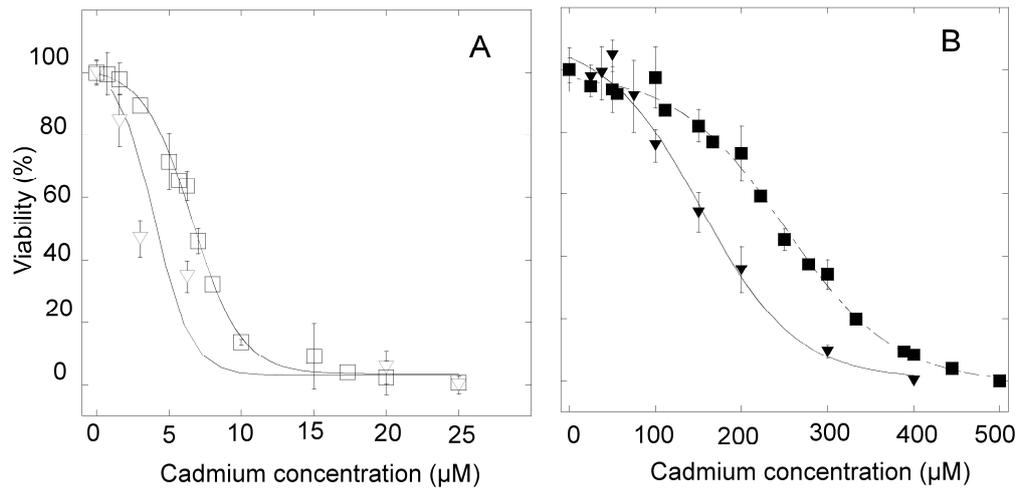

**Figure 8**. **Tyrosine and cadmium toxicity in HeLa and HZR cells. (A)** Parental HeLa cells were stressed with cadmium alone (open squares) or in the presence of 3 mM tyrosine (inversed open triangles) for 24 hours. Eight separate measurements with the MTT assay were used to draw viability curves. **(B)** The same experiments were done with HZR cells and represented with the same filled symbols.